\documentclass{article}
\usepackage{iclr2024_conference,times}
\usepackage{graphicx}
\usepackage{float}
\graphicspath{{./images/}}
\usepackage{booktabs}
\usepackage{authblk}
\begin{document}

\title{ClimateQ\&A : bridging the gap between climate scientists and the general public}
\author{%
Natalia De La Calzada\thanks{natalia.delacalzada@ekimetrics.com} , Théo Alves Da Costa \thanks{theo.alvesdacosta@ekimetrics.com}, Annabelle Blangero\thanks{annabelle.blangero@ekimetrics.com}, Nicolas Chesneau\thanks{nicolas.chesneau@ekimetrics.com}
}

\maketitle

\begin{abstract}
This research paper investigates public views on climate change and biodiversity loss by analyzing questions asked to the ClimateQ\&A platform. ClimateQ\&A is a conversational agent that uses LLMs to respond to queries based on over 14,000 pages of scientific literature from the IPCC and IPBES reports. Launched online in March 2023, the tool has gathered over 30,000 questions, mainly from a French audience. Its chatbot interface allows for the free formulation of questions related to nature*. While its main goal is to make nature science more accessible, it also allows for the collection and analysis of questions and their themes. Unlike traditional surveys involving closed questions, this novel method offers a fresh perspective on individual interrogations about nature. Running NLP clustering algorithms on a sample of 3,425 questions, we find that a significant 25.8\%  inquire about how climate change and biodiversity loss will affect them personally (e.g., where they live or vacation, their consumption habits) and the specific impacts of their actions on nature (e.g., transportation or food choices). This suggests that traditional methods of surveying may not identify all existing knowledge gaps, and that relying solely on IPCC and IPBES reports may not address all individual inquiries about climate and biodiversity, potentially affecting public understanding and action on these issues.

\underline{Note:}we use “nature” as an umbrella term for “climate change” and “biodiversity loss”.
%*Note: we use “nature” as an umbrella term for “climate change” and “biodiversity loss”.

\end{abstract}

% The rest of the document content goes here

\section{Introduction}

Understanding public perception of climate change and biodiversity loss involves assessing a population's awareness and literacy on these issues. Climate change awareness is characterized by concern and belief in the phenomenon \cite{jurkenbeck2021}. On the other hand, climate change literacy, a subset of scientific literacy, denotes the knowledge, skills, and attitudes necessary for effectively addressing climate change. Literacy in climate change is essential for informed decision-making, emissions reduction, and community resilience \cite{lesleyann2018}.

Studies in recent years have sought to quantify public awareness and literacy on nature topics. Whether conducted by academia or international organizations, the predominant method used is closed surveys and questionnaires. Comprehensive examples include the Sustainability Literacy Test (Sulitest) that is aligned with the UN SDGs, and the People’s Climate Vote developed by the UNDP and Oxford University's Department of Sociology \cite{Peoplevote}. Other research groups have examined perspectives from individuals of different age groups and regions, such as responses to climate change in advanced economies \cite{PewResearch}, climate change beliefs in Latin America \cite{Latam}, the influence of socio-demographic characteristics on climate change awareness and risk perception across 119 countries \cite{YaleU}, among many others.

We find that in the cited cases, data is collected through a series of closed questions asked to the participant, for example:

\textit{“"How serious of a threat is global warming to you and your family?’ Response categories included: ‘Not at all serious’, ‘Not very serious’, ‘Somewhat serious’, and ‘Very serious’"} \cite{YaleU}

These studies offer valuable insights into the public'’s perception of climate change. They demonstrate that awareness of climate change significantly differs across geographical regions and demographic groups. While personal experiences and exposure to phenomena like rising temperatures influence awareness, educational attainment emerges as its most significant predictor. Education facilitates access to global discussions on the topic, often conducted in the language of science and global politics \cite{YaleU}. Indeed, the primary sources of information are the reports of the IPCC and IPBES, whose core mandate is to provide policymakers with assessments of climate and biodiversity. While these documents provide a comprehensive understanding of nature phenomena, their format poses several barriers to the general public. These include the use of complex jargon (e.g., uncertainty and confidence levels) and availability exclusively in the UN-official languages, which are spoken as a first or second language by only half of the world'’s population. Additionally, a comprehensive understanding of these documents requires consideration of both the Summary for Policymakers and the Full Reports, which collectively can span thousands of pages for each publication. These barriers, combined with growing fake news and skepticism around these topics, can hinder public awareness and understanding of these critically important issues.

It is in view of these issues that we decided to create ClimateQ\&A (www.climateqa.com). The tool leverages LLMs to interpret and answer nature-related questions based solely on the IPCC and IPBES reports. The architecture of the tool is described in Annex A. Its core functionalities are: a chatbot-like interface in which users can enter questions in any language, settings that allow for the selection of specific reports and language modes (general public, experts and children), and the sources of the information quoted in the answer, with direct access to the documents in the page where the information is found. The tool uses more than 14,500 pages of knowledge from the two institutions, spanning the multiple publications (Full Report, Summary for Policymakers, Special Reports, Cross-Chapter Documents) to deliver a fact-based, comprehensive and understandable answer to the user.

Since its launch in May 2023, the tool has attracted significant attention from the main media outlets in France, namely Libération \cite{Libe}, Le Figaro \cite{Figaro}, Les Echos \cite{Echos} among others; and several authors linked to the IPCC have recommended the use of the tool through social media \cite{Valmasdel}. This important reach has allowed us to build a database of user queries, which we analyze in the next section to understand individual interrogations of climate and biodiversity issues. To our knowledge, this is the first large-scale study that uses this novel open-ended method to derive insights on public perception of nature phenomena.

\section{Method}
We extracted a sample of 7,000 questions posed to ClimateQ\&A in April 2023, one month after the tool's launch. Using open-source NLP algorithms, we categorized these questions into two groups: (1) salient topics, such as climate change, nature, and biodiversity, and (2) intent. From this sample, we excluded questions meeting the following criteria:
\begin{itemize}
\item\- Repeated questions (exact matches), typically those suggested to users on the platform.
\item\- Unclassifiable questions, including those lacking a specific topic, unrelated queries, or those not identified by the NLP algorithm.
\end{itemize}

Subsequently, we conducted clustering based on dense embeddings of the questions (bge-base-en-v1.5). Initial cluster naming was automated using BERTopic. Employing this NLP approach, we categorized 3,425 questions across 130 different clusters representing salient topics. We manually reviewed each cluster to identify recurring categories and sub-categories related to climate and biodiversity science, along with adjacent topics related to economics and society. Finally, we characterized each cluster based on whether the user expressed a general or personal inquiry.

\section{Results}
When conducting a topic analysis, we find that the majority of questions in the sample relate to Climate Change and Greenhouse Gas Emissions (43.1\%). Within this category, questions cover several sub-topics, including the general concept of climate change, its causes, consequences, and potential mitigation or reversal measures. The subsequent largest categories, Earth System (12.1\%), Energy (10.5\%), and Citizens \& Behaviors (5.1\%) are considerably smaller in comparison to climate-related inquiries. Biodiversity-related questions, in particular, are notably less prevalent, accounting for only 2.9\% of the total. This disparity may stem from the tool’s name, ClimateQ\&A, and the tool interface'’s UX emphasis on climate-related queries. Additionally, it could be attributed to the comparatively lower coverage of biodiversity topics in mass media and public discourse compared to climate change. For a complete breakdown of the questions’ topics, refer to Annex B.

During intent analysis, we observe that up to 25.8\% of questions express personal interrogations regarding nature phenomena. The most prevalent sub-topic in the dataset, comprising 8.8\% of the total, pertains to inquiries about climate change impact projections in particular geographic areas, possibly tied to individual user circumstances. Other remarkable categories encompass the impacts on climate change of specific consumption patterns (6.7\%), personal concerns regarding the physical and economic consequences of climate change (5.1\%), skepticism regarding scientific consensus or the ramifications of climate change (0.3\%), and the efficiency of individual actions in mitigating its impacts (2.1\%).

The implications of these findings are twofold. Firstly, they suggest that employing social surveys to gauge public perceptions of nature, where users can freely formulate questions, may help climate scientists identify existing knowledge gaps and pinpoint areas of interest for the general public. This may not be allowed by closed-ended questionnaires, where participants choose from a list of pre-specified options. In our analysis, we find that the public may not only interested in broader concepts of nature science but also in how these phenomena will personally affect them. Secondly, it underscores the need for enhanced efforts to meet this specific demand for climate and biodiversity-related information. Given the IPCC and IPBES's mandate to provide synthesis for policymakers, alternative outreach and dissemination techniques may be necessary, potentially leveraging tools like ClimateQ\&A and tailored communication strategies suited to local contexts.

\section{Discussion and future work}
This paper investigates public perceptions of climate change and biodiversity loss using a dataset comprising 3,500 questions extracted from ClimateQ\&A. This subset corresponds to inquiries made in April 2023, shortly after the tool's launch for public use, and contains sampling bias stemming directly from the French audience of the tool. By February 2024, the tool has gathered over 30,000 questions. A comprehensive analysis of this larger dataset could provide further insights into knowledge gaps and public demand for specific topics. 

An additional interesting avenue for analysis would involve tracking the evolution of questions over time and aligning this evolution with important environmental events (floods, wildfires, heat waves...) or ongoing public debates to assess whether such events influence public interest in related topics. For example, we have identified a significant number of questions associated with large water reservoirs, which may be linked to ongoing related controversies in France in March/April 2023.

Based on our findings, we emphasize that this method of social surveying offers novel perspectives on public perceptions of climate change and biodiversity loss, particularly in terms of awareness and literacy. While traditional survey methods are valuable for establishing baseline attitudes, this approach delves into individuals' personal inquiries about the subject, potentially informing future research and communication strategies at organizational, national, and community levels.

In this context, enhancing the tool on multiple fronts would be beneficial. For instance, enriching the question database with climate change projections specific to countries or regions would enable users to access tailored information directly on the platform, addressing their specific queries about the impacts of climate change in the places where they live or vacation.

Finally, the tool should progressively integrate advancements in the areas of computer vision, allowing it to better integrate and interpret tables and graphics, and generative AI. Some ideas of improvements specific to ClimateQ\&A have already been discussed in the literature \cite{NatureEki}, namely improving the prioritization of information, and the addition of expert judgment.

\newpage
\bibliographystyle{plainnat} % Choose your bibliography style
\bibliography{ICLR}

\appendix

\newpage
\section{Annex A}

ClimateQ\&A is built of several technical algorithmic modules, OpenAI GPT 3.5 API being the last one (the generation of an answer). The core of Climate Q\&A consists of three steps, summarized in Figure 1: creation of a structured dataset, question enrichment and sourcing, and generation \& display in an interface.

\begin{figure}[H]
    \centering
    \includegraphics[width=1\textwidth]{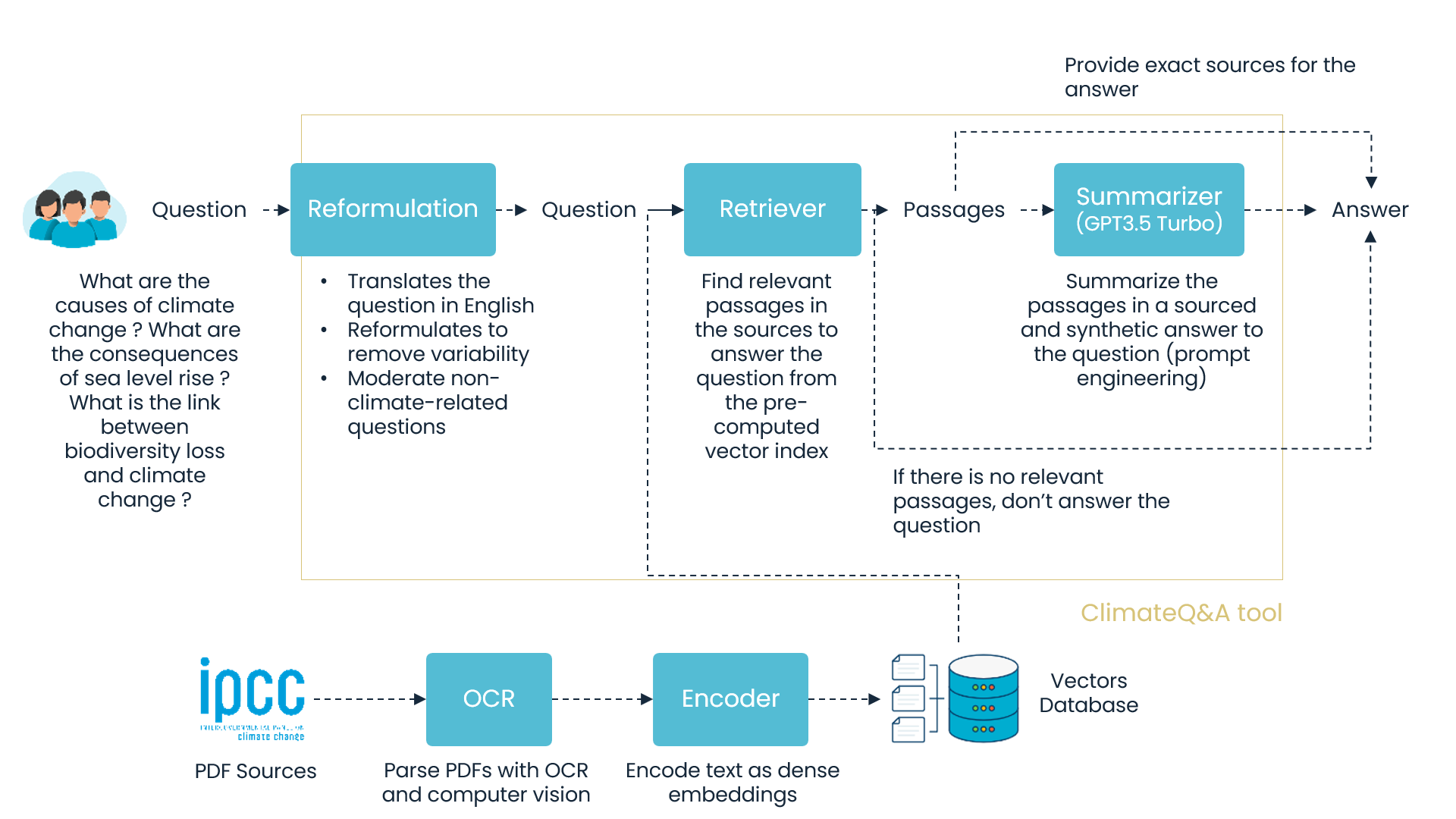}
    \caption{Climate Q\&A global architecture}
    \label{fig:enter-label}
\end{figure}

\subsection{Step 1: Creation of a structured dataset from a set of heterogeneous documents}
The first step parses a heterogeneous set of documents (IPCC and IPBES reports) and extracts information, which is stored in a structured database.
The algorithm uses OCR techniques for all the documents stored in any given folder. The document structure is preserved, meaning that paragraphs are linked to section titles, figures are linked to the passages that cite them, etc. A structured database is thus created, containing all the information extracted from the documents and the relationships between the different entries in the database. These entries are then represented as a vector of finite dimension. In ClimateQ\&A, we use SentenceBert, but any type of representation is possible. This will serve as the basis for a quick search in the database for step 2. 

\subsection{Step 2: Enrichment of the user's LLM query}
In the second step, the user's query is reformulated and enriched by searching for similar passages in the structured database. The relevant question and passages are encapsulated in a prompt, which also serves to limit the scope of acceptable answers for ClimateQ\&A.
Within the step of enriching the user's query, the goal is to create a query on an LLM from the question, whose answer will be formulated from the documents of interest (the structured database from step 1). First, the user's question is reformulated in a more intelligible way for a LLM by asking the model to do it. Then, the new question is compared to all entries in the database using the Faiss algorithm developed by Facebook (Meta), known for its execution speed on data corpora up to billions of entries. The most relevant entries, meaning those with the most significant similarity to the question, are selected, filtered by a thematic classification model (in this case, if the entry discusses climate) and encapsulated in a prompt, necessary for querying an LLM (in our case, GPT-3.5 Turbo). The prompt contains other information, such as not going beyond the scope of its knowledge and formulating its answer based on the selected entries. The prompt created is then used to query an LLM.

\subsection{Step 3 : Displaying the results}
The query result (i.e.; the question) is displayed on an interface. The answer is generated using the 10 most relevant references found in the corpus of documents. The sources are displayed alongside the answer, so that the user can verify the information and extract it for a report if necessary.
The display is based on two parts: the LLM's answer with the notes serving as a reference, as well as the sources used to formulate the answer. For ClimateQ\&A, we use the OpenAI Azure environment to send the request and get the response. Just like the ClimateQ\&A interface, the user can continue their search by asking several questions in a row like a real chatbot.

\newpage
\section{Annex B}

Table 1 presents the results of the breakdown of the questions by topic. 

\begin{table}[htbp]
    \centering
    \begin{tabular}{lrr}
        \toprule
        Topics & Number of questions & Share of total \\
        \midrule
        Climate Change \& Greenhouse gas emissions & 1477 & 43.1\% \\
        Earth system & 416 & 12.1\% \\
        Energy & 358 & 10.5\% \\
        Citizens \& behavior & 173 & 5.1\% \\
        Economics & 171 & 5.0\% \\
        IPCC-related questions & 170 & 5.0\% \\
        Companies \& industries & 153 & 4.5\% \\
        Food \& Agriculture & 104 & 3.0\% \\
        Feelings & 101 & 2.9\% \\
        Biodiversity & 101 & 2.9\% \\
        Politics \& Policies & 70 & 2.0\% \\
        Circular economy & 67 & 2.0\% \\
        Technologies & 35 & 1.0\% \\
        Social issues & 29 & 0.8\% \\
        \midrule
        Total & 3425 & 100\% \\
        \bottomrule
    \end{tabular}
        \caption{Breakdown of questions asked to ClimateQ\&A by topic}

\end{table}

The topics in Table 1 are defined as follows.

\underline{Climate change and greenhouse gas emissions:} Causes of climate change, how to act against climate change, what is climate change, climate risks and impacts, carbon emissions, temperature projections, evidence of climate change, impacts of climate change in France...

\underline{Earth system:} Sea level rise, ocean currents, water, forests, permafrost, icecap melt, glaciers, planetary boundaries.

\underline{Energy:} Electric vehicles, renewable energies, nuclear energy, carbon capture \& storage, hydrogen, fossil fuels, biochar \& bioenergies, batteries.

\underline{Citizens \& behavior:} Individual action, food \& transportation habits, effectiveness of individual action.

\underline{Economics:} Impact of different sectors of activity, welath \& carbon footprint, degrowth, green economy, financial system.

\underline{IPCC-related questions:} Questions about the reports \& authors, the main conclusions, the existence of a consensus.

\underline{Companies \& industries:} Sustainability \& CSR, role of companies in the transition.

\underline{Food \& Agriculture:} Agricultural practices, animal products, risks to specific agricultural products.

\underline{Feelings:} Death \& survival, climate sceptics.

\underline{Biodiversity:} Link between climate change \& biodiversity loss, impacts of biodiversity loss, mass extinctions.

\underline{Politics \& Policies:}International agreements, political parties \& ideologies.

\underline{Circular economy:} Waste management, recycling \& waste.

\underline{Technologies:} Digital technologies and AI.

\underline{Social issues:} Gender issues and social justice.

\begin{figure}[H]
    \centering
    \includegraphics[width=1\textwidth]{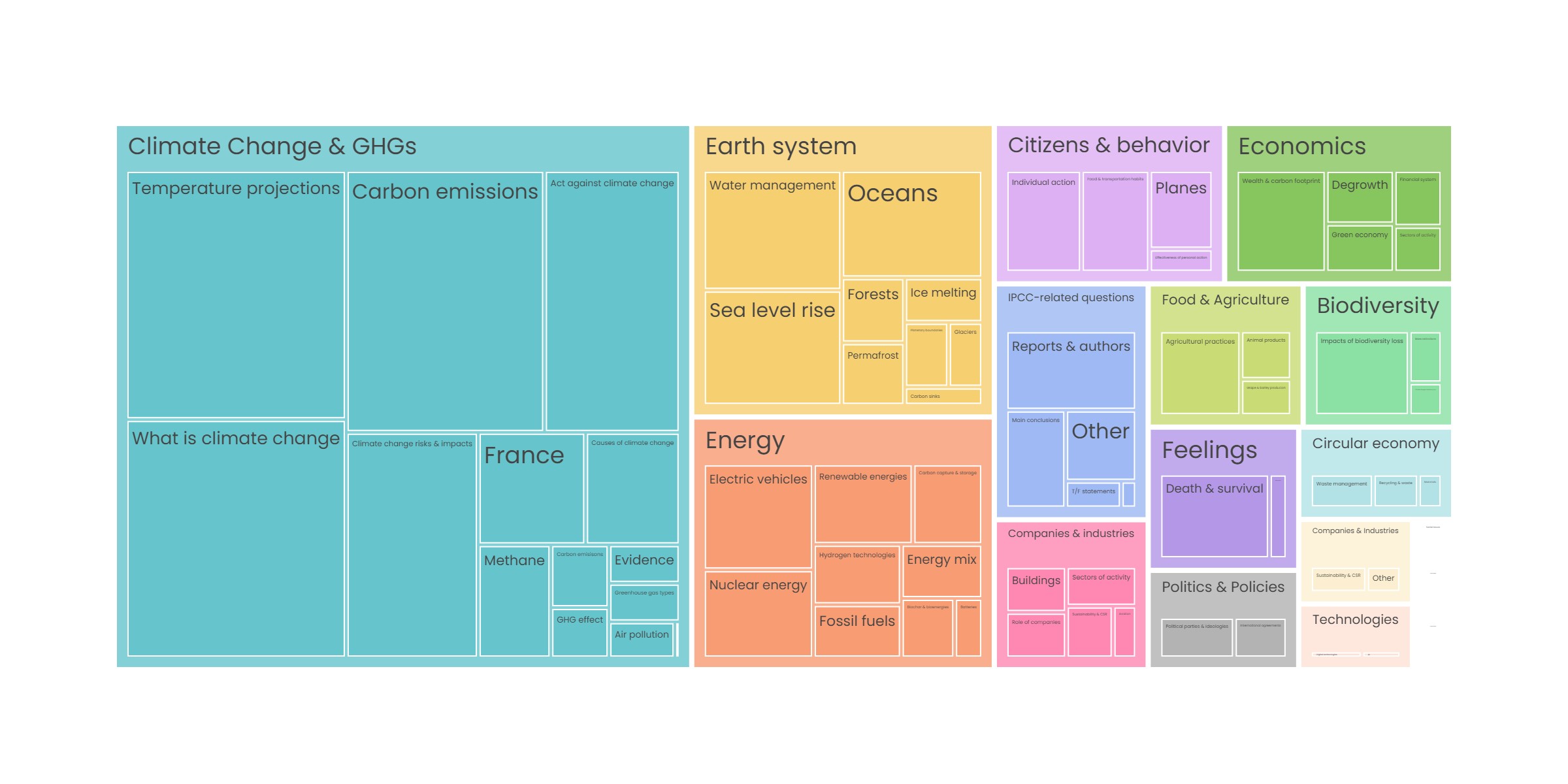}
    \caption{Illustration of the breakdown of the questions by topic and subtopic}
    \label{fig:enter-label}
\end{figure}

\end{document}